\documentclass[aps,pra,twocolumn,superscriptaddress]{revtex4-1}

\usepackage{graphicx}
\usepackage{hyperref}
\usepackage{amsmath}
\usepackage{amssymb}
\usepackage{epstopdf}
\usepackage{multirow}
\usepackage{threeparttable}
\usepackage[usenames]{color}

\begin{document}

\title{Single-photon time-stretch infrared spectroscopy}
\author{Ben Sun}
\affiliation{State Key Laboratory of Precision Spectroscopy, East China Normal University, Shanghai 200062, China}

\author{Kun Huang}
\email{khuang@lps.ecnu.edu.cn}
\affiliation{State Key Laboratory of Precision Spectroscopy, East China Normal University, Shanghai 200062, China}
\affiliation{Chongqing Key Laboratory of Precision Optics, Chongqing Institute of East China Normal University, Chongqing 401121, China}
\affiliation{Collaborative Innovation Center of Extreme Optics, Shanxi University, Taiyuan, Shanxi 030006, China}

\author{Huijie Ma}
\affiliation{State Key Laboratory of Precision Spectroscopy, East China Normal University, Shanghai 200062, China}

\author{Jianan Fang}
\affiliation{State Key Laboratory of Precision Spectroscopy, East China Normal University, Shanghai 200062, China}

\author{Tingting Zheng}
\affiliation{State Key Laboratory of Precision Spectroscopy, East China Normal University, Shanghai 200062, China}

\author{Yongyuan Chu}
\affiliation{Key Laboratory of Specialty Fiber Optics and Optical Access Networks, Shanghai University, Shanghai 200444, China}

\author{Hairun Guo}
\affiliation{Key Laboratory of Specialty Fiber Optics and Optical Access Networks, Shanghai University, Shanghai 200444, China}

\author{Yan Liang}
\affiliation{School of Optical Electrical and Computer Engineering, University of Shanghai for Science and Technology, Shanghai 200093, China}

\author{E Wu}
\affiliation{State Key Laboratory of Precision Spectroscopy, East China Normal University, Shanghai 200062, China}
\affiliation{Chongqing Key Laboratory of Precision Optics, Chongqing Institute of East China Normal University, Chongqing 401121, China}

\author{Ming Yan}
\affiliation{State Key Laboratory of Precision Spectroscopy, East China Normal University, Shanghai 200062, China}
\affiliation{Chongqing Key Laboratory of Precision Optics, Chongqing Institute of East China Normal University, Chongqing 401121, China}

\author{Heping Zeng}
\email{hpzeng@phy.ecnu.edu.cn}
\affiliation{State Key Laboratory of Precision Spectroscopy, East China Normal University, Shanghai 200062, China}
\affiliation{Chongqing Key Laboratory of Precision Optics, Chongqing Institute of East China Normal University, Chongqing 401121, China}
\affiliation{Shanghai Research Center for Quantum Sciences, Shanghai 201315, China}
\affiliation{Chongqing Institute for Brain and Intelligence, Guangyang Bay Laboratory, Chongqing, 400064, China}

\begin{abstract}
Sensitive mid-infrared (MIR) spectroscopy is highly demanded in various fields ranging from industrial inspection, biomedical diagnosis to astronomical observation. However, the detection sensitivity of conventional MIR spectrometers has been severely limited by excessive noises for existing infrared sensors, which hinders widespread use in photon-scarce scenarios. Here, we devise and implement a broadband MIR single-photon time-stretch spectrometer based on high-fidelity spectral upconversion and time-correlated coincidence counting. Specifically, a nanophotonic supercontinuum illumination covering 2.4-4.2 $\mu$m is nonlinearly converted to the near-infrared band, where low-loss single-mode fiber and high-performance silicon detector can be leveraged to facilitate dispersive operation and sensitive detection, respectively. The arrival time for the dispersed upconversion photons is precisely registered with a low-timing-jitter photon counter, which enables us to obtain a high spectral resolution about 0.5 cm$^{-1}$ under a low-light-level illumination down to 0.14 photons/nm/pulse. In comparison to previous MIR upconversion spectrometers, the presented time-stretch architecture favors single-pixel simplicity and high-throughput acquisition for the single-photon spectral measurement. The achieved MIR spectroscopic features of broadband spectral coverage, sub-wavenumber resolution, single-photon sensitivity, and room-temperature operation would stimulate immediate applications in material and life sciences.
\end{abstract}

\maketitle

\section{Introduction}
Mid-infrared (MIR) spectroscopy has become an indispensable tool to probe distinct rotational and vibrational modes of specific molecules, which is extensively used to identify chemical species and functional groups in material, biology and medicine fields, among others \cite{Vodopyanov2020Book, Fernandez2005NB, Cheng2015Science}. Nowadays, there is a great impulse to develop sensitive MIR spectroscopy to address the increasing need in a variety of photon-scarce applications, such as trace detection \cite{Haas2016ARAC}, remote sensing \cite{Fang2020NC}, environmental monitoring \cite{Coburn2018Optica}, biological examination \cite{Shi2020NM}, and heritage conservation \cite{Daffara2018OLE}. In these scenarios, the collected photon flux is stringently limited due to long operation distance, weak backward scattering, or minimized radiation exposure. Consequently, high-contrast spectral measurements imperatively call for highly-sensitive MIR detection that challenges conventional infrared sensors based on narrow-bandgap semiconductors or thermal-type bolometers \cite{Razeghi2014PR, Wang2019Small, Liu2021LSA}. Indeed, typical MIR detectors are facing limitations in high dark current, slow response time, and low operation temperature, thus struggling to implement sensitive and fast MIR spectroscopy \cite{Israelsen2019LSA, Zheng2023LPR}. In addition, scanning-free MIR spectrograph usually requires focal plane arrays (FPAs), where the spectral resolution is severely restricted by the low pixel density \cite{Razeghi2014PR}. Recently, superconducting nanowire detectors have been integrated to implement broadband single-photon spectrometers \cite{Cheng2019NC, Kong2021NL, Xiao2022ACS}, albeit with limited operation wavelengths below 2 $\mu$m and additional system complexity in cryogenic cooling. Therefore, it remains a long-sought-after quest to realize high-resolution and high-sensitivity MIR spectroscopy at room temperature over a wide spectral range.

In this context, an indirect strategy for MIR spectroscopy has been proposed to circumvent the aforementioned limitations of arrayed MIR detectors, where an intermediate process is used to transfer the MIR information into the visible or near-infrared bands for accessing more mature large-bandgap semiconductor technology \cite{Barh2019AOP}. So far, there have been various approaches to implement the involved frequency transduction step. One common way is based on frequency upconversion of the MIR photons through a high-efficiency and low-noise parametric nonlinear conversion, which allows for demonstrating broadband MIR spectrometers at single-photon sensitivity \cite{Dam2012NP, Mancinelli2017NC, Cai2022PR, Huang2022NC} or kHz-above frame rates \cite{Zheng2023LPR, Wolf2017OE, Petersen2018OL}. Another promising approach resorts to quantum interferometry of spectrally-entangled photons from spontaneous parametric down conversion (SPDC) \cite{Kalashnikov2016NP, Lindner21OE}, which features intrinsic single-photon-level sensitivity. In both methods, the involved spectral analysis is often conducted in the spatial domain based on a monochromator \cite{Mancinelli2017NC, Cai2022PR} or a spectrograph \cite{Zheng2023LPR, Wolf2017OE}. The former typically requires stringent spectral filtering to obtain a high resolution based on a rotational grating and a narrow slit. The resulting low light throughput may result in a long acquisition time especially at photon-starving circumstances. Alternatively, the latter allows for parallel spectral recording based on a multi-pixel detector. However, the required single-photon array detector is technically challenging and expensive for high-definition pixel formats. Furthermore, the number of spectrally resolved elements is determined by the pixel number, which inevitably leads to a trade-off between spectral coverage and spectral resolution \cite{Zheng2023LPR, Wolf2017OE, Rodrigo2021LPR}.

In parallel, optical time-stretch technique has emerged to provide an effective solution to improve the spectroscopic performance, where the spectral components are temporally spread according to the group velocity dispersion (GVD) within a highly dispersive propagation channel \cite{Solli2008NP, Goda2013NP, Zhou2022LPR}. The underlying wavelength-time correspondence facilitates high-throughput and high-resolution spectral measurement via a high-bandwidth single-pixel detector in the time domain \cite{Zhou2022LPR}. Such time-stretch spectroscopy is initially devised to realize single-shot measurement for capturing non-repetitive or transient events \cite{Goda2013NP}, which has now been extended to the single-photon regime by combing the time-correlated single photon counting (TCSPC) technique. For instance, the single-photon time-stretch concept has been adopted in photon-pair characterization \cite{Avenhaus2009OL, Eckstein2014LPR, Davis2017OE, Yang2020OE}, photonic waveform reconstruction \cite{Crockett2022LPR}, quantum microwave photonics \cite{Yang2022SB}, single-pixel fluorescence analysis \cite{Tiedeck2022ACS}, and lightweight Raman spectroscope \cite{Meng2015PNAS}. However, all these demonstrations are restricted in the near-infrared region. The implementation of sensitive MIR time-stretch spectroscopy is hindered by the deficiencies on low-loss dispersive media and high-sensitivity photon detectors \cite{Kawai2020CP}. Very recently, upconversion time-stretch infrared spectroscopy has been explored, yet the detection sensitivity is far away from the single-photon level \cite{Wen2023LPR, Hashimoto2023LSA}. Intrinsically, the adopted conversion process based on different-frequency generation (DFG) suffers from severe spontaneous parametric noises. To date, it has still been a challenging goal to realize single-photon time-stretch MIR spectroscopy.

Here, we have devised and implemented an ultra-sensitive MIR time-stretch spectroscopy based on chirped-pumping sum frequency generation (SFG). The MIR illumination is provided by a turn-key and compact broadband supercontinuum of 2.4-4.2 $\mu$m based on a nanophotonic silicon nitride (Si$_3$N$_4$) waveguide. The involved coherent nonlinear conversion facilitates a high-fidelity spectral translation of the incident MIR field into the near-infrared region in a high-efficiency and low-noise fashion. The upconverted photons are then temporally dispersed by using low-loss single-mode fibers, and detected by a low-timing-jitter silicon photon counter. The coincidence registration for relative time between pump pulses and the detected photons allows for reconstructing the time-stretch infrared spectrum at sub-wavenumber spectral resolution and single-photon detection sensitivity. In contrast to the spatial counterpart, the presented time-domain spectroscopy enables us to implement high-throughput acquisition for parallel spectral components with a single-element detector, which might promote a variety of low-light-level applications in trace inspection, remote sensing, and biomedical examination.
 
\section{Basic principle}
\begin{figure*}[t!]
\includegraphics[width=0.85\textwidth]{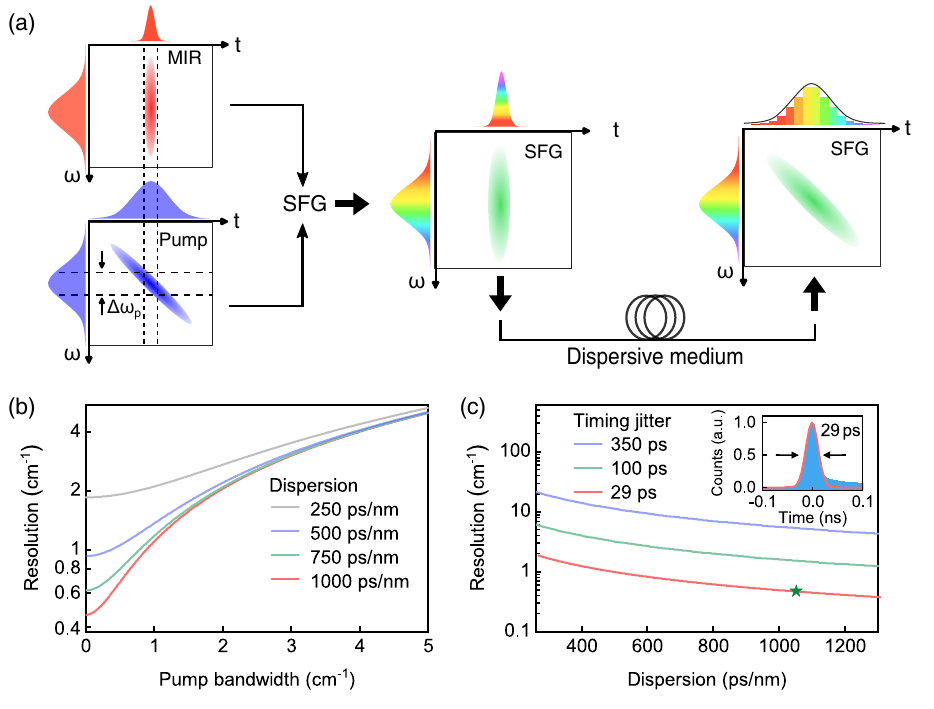}
\caption{Concept and performance of the MIR time-stretch single-photon upconversion spectroscopy. (a) Schematic diagram for the involved key steps. A chirped-pump SFG process favors a high-fidelity spectral transduction, where only a narrow portion of the pump spectrum contributes to the nonlinear optical mixing. The upconversion field is then sent into a dispersive medium for implementing the time-stretch operation based on group velocity dispersion. The temporal profile of the broadened SFG pulse can be reconstructed by using the time-correlated photon counting technique. Finally, the measured coincidence histogram allows us to reveal the infrared spectrum. (b) Theoretical resolution for the time-stretch spectrometer as a function of the pump bandwidth in the presence of various dispersions. (c) Spectral resolution varies as the increased dispersion in the case of different timing jitters for the single-photon detector. The star symbol in dark green indicates the expected resolution for our experimental settings. Inset shows the measured timing-jitter trace of the used detector. The full width at half maximum is inferred to be 29 ps from the Gaussian fitting.}
\label{fig1}
\end{figure*}

The core of the proposed MIR upconversion time-stretch spectroscopy lies in spectro-temporal manipulation of the involved optical fields as illustrated in Fig. \ref{fig1}. The whole process consists of three major steps: broadband frequency upconversion, time-stretching operation, and coincidence photon counting. The nonlinear conversion is facilitated in pulsed pumping configuration, where the high peak power and the ultrashort pulse duration favor efficiency improvement and noise suppression \cite{Zheng2023LPR}. As a result of the three-wave mixing process in SFG, one prominent concern for the upconversion spectrometer is the degradation of spectral resolution due to the broad pump bandwidth. To circumvent the limitation, a chirped pulse pump is engineered to implement a high-fidelity spectral transduction \cite{Kubarych2005OL}. In this case, the pump pulse is typically extended to tens of picoseconds, which is much longer than the femtosecond MIR pulse \cite{Shirai2015PRAppl}. Consequently, the MIR pulse temporally overlaps only with a small portion of frequencies inside the chirped pulse, hence resulting in a narrow-pumping capability. Such a chirped pulse upconversion has been applied to MIR absorption spectroscopy \cite{Kubarych2005OL, Zhu2012OE} or pump-probe spectroscopy \cite{Shirai2015PRAppl}. In these demonstrations, the spectral information is resolved in the spatial domain. The achieved spectral resolution is typically at several cm$^{-1}$, which is limited by the performance of dispersive gratings or array detectors.

In this context, time-domain spectroscopic measurement is conducted with an aim to achieve a high spectral resolution, without the loss of desirable features of wide spectral coverage and scanning-free operation \cite{Solli2008NP, Goda2013NP}. Here, the upconverted pulse is sent into a dispersive medium to transform the temporal envelope into its spectrum via the GVD effect. The wavelength-to-time mapping allows one to directly acquire the spectrum in the time domain with a single-element detector. The time-stretch paradigm has been intensively investigated in the visible or near-infrared regime \cite{Goda2013NP}. However, MIR time-stretch spectroscopy has long remained elusive due to the lack of  low-loss dispersive media and large-bandwidth detectors. For instance, the loss of silica fibers at near-infrared wavelengths is below 1 dB/km, while MIR fluoride fibers suffer from much higher losses about 100 dB/km \cite{Jackson2012NP}. Moreover, the fabrication process for the MIR fibers is complicated and expensive. Meanwhile, sensitive and fast detection in the MIR regime is challenging. The bandwidth of cutting-edge commercial MIR detectors is usually at the GHz level \cite{Razeghi2014PR, Kawai2020CP},  which is far way from the typical requirements about 20-50 GHz to perform high-precision time-domain sampling. In our scheme, the involved frequency upconversion provides an effective solution to mitigate the above restrictions in MIR performances. Particularly, the time-stretched SFG pulse is registered by using a TCSPC module. The measured coincidence count histogram of the relative time between the pump trigger and temporally dispersed photon makes it possible to implement the MIR single-photon time-stretch spectrometer. Notably, the involved signal processing bandwidth is mainly determined by the timing jitter of the single-photon detectors. Therefore, the single-photon detection not only significantly enhances the measurement sensitivity, but also overcomes the stringent bandwidth limitation for the electronic transmission and processing in conventional time-stretch systems \cite{Yang2022SB}.

\begin{figure*}[t!]
\includegraphics[width=0.85\textwidth]{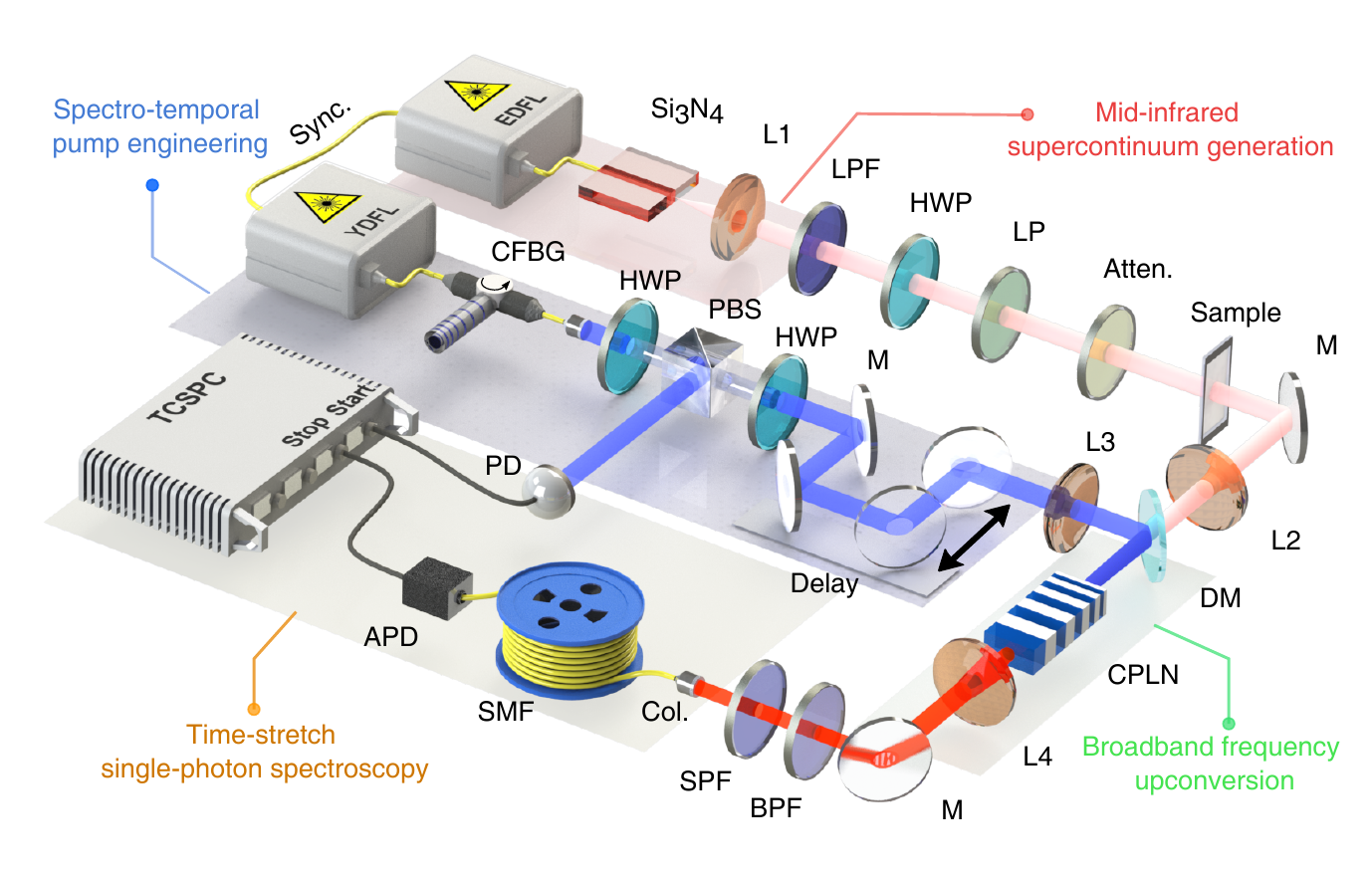}
\caption{Experimental setup. A broadband MIR supercontinuum source is prepared with a nanophotonic Si$_{3}$N$_{4}$ waveguide that is driven by a femtosecond Er-doped fiber laser (EDFL). The generated MIR pulse is steered through a long-pass filter (LPF) to extract the spectral portion above 2.4 $\mu$m. The beam polarization is manipulated by a half-wave plate (HWP), and enforced by a vertically-orientated linear polarizer (LP). The MIR light illuminates the targeted sample after an optical attenuator. In parallel, the pump pulse originates from a synchronous Yb-doped fiber laser (YDFL), which passes through a chirped fiber Bragg grating (CFBG). The pump and infrared pulses are spatially combined with a dichroic mirror (DM), while the temporal overlap is optimized by a delay line. The mixed beam is sent into a chirped-poling lithium niobate (CPLN) crystal to perform the broadband frequency upconversion. After a band-pass filter (BPF) and a short-pass filter (SPF), the upconverted photons are coupled via a collimator into a long haul of single-mode fiber (SMF). The dispersed waveform at the single-photon level is measured by the time-correlated single photon counting (TCSPC) technique. The start channel is triggered by the pump pulse, and the stop channel is fired by detected events from an avalanche photodiode (APD). The resulting coincidence histogram enables us to reconstruct the infrared spectrum at high detection sensitivity and high spectral resolution.}
\label{fig2}
\end{figure*}

\begin{figure*}[t!]
\includegraphics[width=0.7\textwidth]{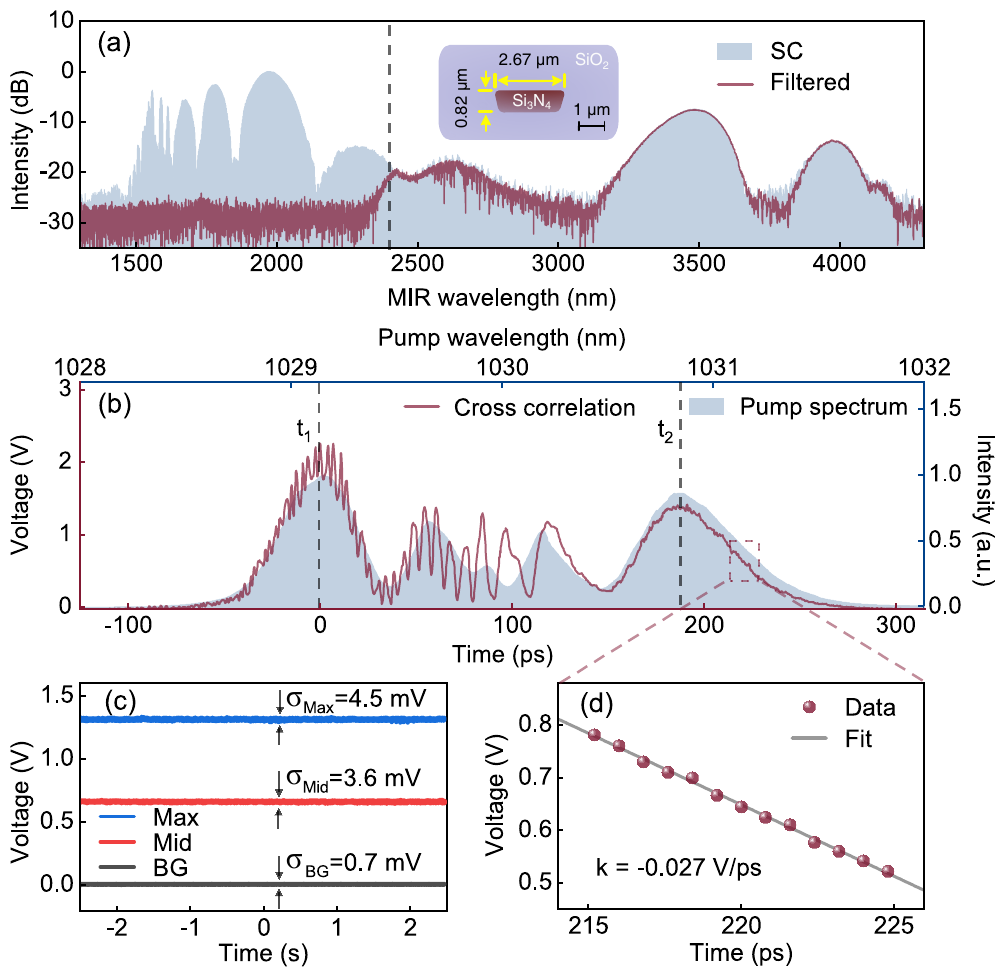}
\caption{Spectro-temporal characterization of signal and pump pulses. (a) Measured optical spectrum (shaded area) for the generated MIR supercontinuum. The solid red line indicates the filtered spectrum after a long-pass filter with a cut-off wavelength at 2.4 $\mu$m. Inset presents the geometric sizes for the Si$_3$N$_4$ waveguide. (b) Measured cross-correlation trace between the dispersed pump pulse and MIR signal pulse. The temporal trace is analogy to the pump spectral profile (shaded area) with a conversion factor about 100 ps/nm. Dashed lines indicate two representative positions for the relative time delay, which will lead to a spectral shift for the upconverted field. (c) Measured cross-correlation intensities at three positions at the trailing edge, which are used to infer the relative timing jitter between the dual-color pulses. (d) Zoom-in illustration for the linear region on the trailing trace. The measured intensity variation can be translated to the timing jitter via the measured slope.}
\label{fig3}
\end{figure*}

To investigate the spectral resolution of the upconversion time-stretch spectrometer, we assume a single-frequency MIR light as the input. After the SFG, the upconversion spectrum is broadened to the pump linewidth $\Delta \omega_p$. The propagation of the SFG pulse through a dispersive element is governed by the nonlinear Schr\"{o}dinger equation \cite{Goda2013NP}, which leads to
\begin{equation}
|E(L,T)|^2 =  \frac{2}{\pi \beta L} e^{-\alpha L} |\tilde{E}(0, \frac{T}{\beta L})|^2 \ ,
\label{eq1}
\end{equation}
where $E$ is the field amplitude of the optical pulse, $\tilde{E}$ denotes the Fourier transform of the signal, $\alpha$ is the absorption coefficient, $L$ is the propagation distance, $\beta$ is the second-order dispersion coefficient, and $T$ is the relative time in the moving frame of the propagation pulse. Also, the stationary phase approximation is assumed at the presence of a large GVD, which is analogous to the requirement for the far-field approximation in paraxial diffraction \cite{Goda2013NP}. In this case, there is a one-to-one mapping between the optical frequency and time, according to $T(\omega) = \beta L (\omega - \omega_0)$. For a single-photon detector with a timing jitter $\tau$, the measured spectral uncertainty $\Delta \omega_d$ is translated to be $\tau/(\beta L)$. By combining the two limiting factors, the overall spectral resolution is found to be $ \Delta \omega = \sqrt{\Delta \omega_p^2 +  \Delta \omega_d^2}$. Figure \ref{fig1}(b) presents the calculated spectral resolution as a function of the pump bandwidth. In the extreme case of $\Delta \omega_p \to 0$, a larger dispersion plays a more prominent role in the resolution improvement. With experimental parameters of $\beta$ =-130.7 ps/nm/km, $L$=8 km, and $\tau$=29 ps, the spectral resolution is deduced to be about 0.44 cm$^{-1}$ as depicted in Fig. \ref{fig1}(c). Notably, under a sufficiently large dispersion, the resolution will be only determined by the pump bandwidth \cite{Yang2020OE}.

\section{Experimental results}
Figure \ref{fig2} illustrates the experimental setup for the MIR time-stretch single-photon upconversion spectrometer, which consists of light source preparation, broadband frequency upconversion, and time-stretch spectroscopy. The involved light sources are originated from a passively synchronized fiber laser system that consists of an Er-doped fiber laser (EDFL, LangyanTech, ErFemto Elite) and an Yb-doped fiber laser (YDFL, LangyanTech, YbPico Elite) in a master-slave configuration. The repetition rates for the EDFL and YDFL are 97 MHz and 19.4 MHz, respectively. The lower repetition rate for the YDFL not only improves the synchronous stability due to a longer interaction length in the salve cavity, but also favors to obtain a higher peak power due to the smaller pulse duty cycle. More details about the implementation for the all-optical synchronization could be found in \cite{Zheng2023LPR}. The average power of the EDFL at 1550 nm is boosted to about 230 mW before being coupled into a nanophotonic Si$_3$N$_4$ waveguide (fabricated by LIGENTECH) with the help of a fiber tip. The height and width of the waveguide section are 0.82 and 2.67 $\mu$m, which are critical to tailor the overall dispersion of the waveguide \cite{Guo2018NP}. With a 5-mm-length waveguide, we obtain an octave-spanning supercontinuum from 1.5 to 4.2 $\mu$m, as  presented in Fig. \ref{fig3}(a). The MIR pulse duration is inferred to about 128 fs according to numerical simulation \cite{Zheng2023LPR}. The supercontinuum light is filtered by a long-pass filter to extract the portion above 2.4 $\mu$m. The generated ultrafast MIR broadband source is essential to implement the subsequent time-stretch spectroscopy, which contrasts to continuous-wave thermal emitters \cite{Dam2012NP}. 

In parallel, the YDFL at 1030 nm is amplified to serve as the pump source. The pump pulse passes through a chirped fiber Bragg grating (CFBG, AFR, PSCG-1030-10.0-1) specified with a dispersion about 100 ps/nm. The resultant pump beam with a broadened pulse duration is spatially and temporally overlapped with the MIR signal in a chirped-poling lithium niobate (CPLN, CTL Photonics, M1900-2400Chirp) crystal to perform a broadband frequency upconversion. The poling period of the nonlinear crystal ramps from 19 to 24 $\mu$m along the length of 20 mm, which allows an acceptance window up to 5 $\mu$m \cite{Zheng2023LPR}. The crystal is operated at room temperature, thus resulting in a simple and robust operation. Since the MIR pulse is much shorter than the pump, the measured cross-correlation trace is similar to the pump spectral profile as shown in Fig. \ref{fig3}(b). The two spectral peaks denoted with dashed lines have a spectral separation of 1.74 nm, which corresponds to a temporal delay of 186 ps. The effective pump bandwidth about 0.011 cm$^{-1}$ is deduced from the spectral portion overlapped with the MIR pulse, thus ensuring a high-fidelity nonlinear spectral transduction. The maximum conversion efficiency is about 0.4\% at a pump power of 0.9 W, which is lower than our previous value due to the longer chirped pump pulse \cite{Zheng2023LPR}. The efficiency can be further increased by boosting the pump power.

\begin{figure}[b!]
\includegraphics[width=0.85 \columnwidth]{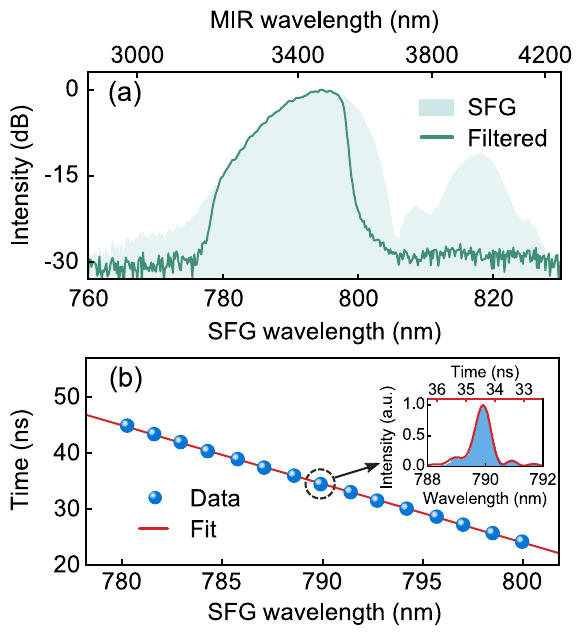}
\caption{Characterization of the upconversion field. (a) Recorded optical spectrum (shaded area) for the upconverted signal, and the green line indicates the filtered spectrum after a band-pass filter. (b) Measured time delay as a function of the center wavelength of the acousto-optic tunable filter. The solid line denotes a linear fit. Inset shows an example of narrow-band filtered spectrum used for calibration. The red line denotes the corresponding coincidence histogram.}
\label{fig4}
\end{figure}

The relative timing jitter $\delta t$ between the pump and signal pulses is characterized by recording the SFG intensity variations within the linear trailing region as shown in Fig. \ref{fig3}(d). The slope is inferred to be $k$=-27 mV/ps. Figure \ref{fig3}(c) presents the standard deviations of the recorded waveforms as $\sigma_\text{Max}$, $\sigma_\text{Mid}$, $\sigma_\text{BG}$, which correspond to the top, middle and bottom positions on the trailing trace. Specifically, $\sigma_\text{BG}$ is ascribed to the detector noise, while $\sigma_\text{Max}$ is determined by the power fluctuation of the source itself \cite{Zheng2023LPR}. As a result, the timing jitter is deduced by $\delta t = [(\sigma_\text{Mid} - \sigma_\text{BG}) - (\sigma_\text{Max} - \sigma_\text{BG})/2]/|k|$=37 fs, which is negligible in subsequent time-stamp measurements. 

Then, the upconverted light is coupled into a section of single-mode fiber (SMF, YOFC, CS780-125-14/250) with a length of 8 km. The fiber has a mode field diameter of 4.5 $\mu$m, a propagation loss around 3.1 dB/km, and a GVD about -131 ps/nm/km at 800 nm. The temporally dispersed photons are registered by an avalanche photodiode (APD, MPD, PD-050-CTD-FC) with a detection efficiency about 13\% at 790 nm. The timing jitter is measured to about 29 ps in full width at half maximum, as shown in the inset of Fig. \ref{fig1}. Based on a TCSPC module (Qutools, quTAG), the time-stretch profile can be revealed from the histogram for the relative time between pump pulse triggers and SFG photons \cite{Avenhaus2009OL, Davis2017OE}. In the coincidence-counting architecture, the time-domain sampling precision is determined by the timing jitter of involved measurements, which greatly relaxes the bandwidth requirement for the involved electronic transmission and processing \cite{Yang2022SB}.

Now we turn to calibrate the MIR time-stretch upconversion spectrometer. Figure \ref{fig4}(a) presents the upconversion spectrum of the MIR supercontinuum illumination. A band-pass filter is used to select the wavelength components from 780 to 800 nm. The filtered bandwidth is selected to adapt the pulse interval of 51.5 ns for the light source in our experiment. The temporally dispersed photons can be ensured to locate within the unambiguous temporal range during the coincidence counting measurement, which is especially important in the case of a large GVD. Notably, the filtration is not mandatory when the laser operates at a lower repetition rate. This can be practically realized by using a pulse picker.

\begin{figure}[t!]
\includegraphics[width=0.85\columnwidth]{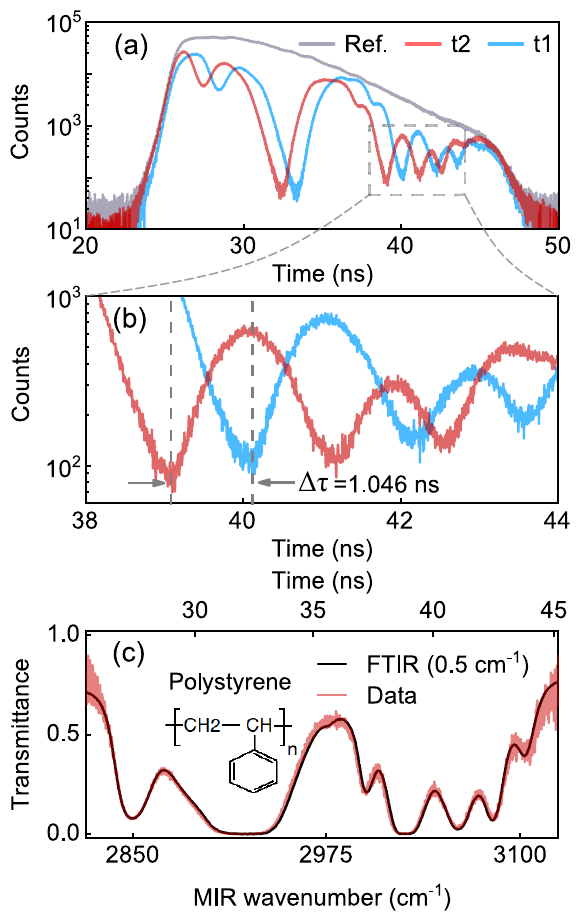}
\caption{High-resolution MIR time-stretch spectroscopy. (a) Recorded histograms for the coincidence photon counts at the presence and absence of the polystyrene film sample. Two pump delays at $t_1$ and $t_2$ are selected for comparison as denoted in Fig. \ref{fig3}(b), which lead to a temporal shift for the two measured histograms. (b) Zoom-in illustration indicates a temporal shift of 1.046 ns, which is expected from the variation of the pump central wavelength. (c) Inferred MIR absorption spectrum based on the wavelength-time correspondence, which agrees well with the reference given by a measured FTIR trace at a spectral resolution of 0.5 cm$^{-1}$.}
\label{fig5}
\end{figure}

The mapping between the time delay and the central wavelength is investigated by inserting an acousto-optic tunable filter (AOTF). The AOTF bandwidth is about 0.9 nm, and the wavelength is electronically tuned from 780 to 800 nm. At each wavelength, the corresponding delay is identified from the peak of the recorded histogram. As shown in Fig. \ref{fig4}(b), the measured data indicates a linear dependence with a slope of -1.0455 ns/nm. A good fitting quality is verified by a coefficient of determination of $R^2$=0.99974. The mapping factor allows one to convert the temporal histogram into the spectral waveform in the time-stretch spectroscopy.

Figure \ref{fig5}(a) shows the measured coincidence-counting histogram for the time delay of temporally dispersed SFG photons relative to the pump pulse. The reference histogram corresponds to the upconversion spectrum given in Fig. \ref{fig4}(a). It can been seen that a longer-wavelength component travels at a faster group velocity due to the normal dispersion of the single-mode fiber. At the presence of a polystyrene film with a thickness of 50 $\mu$m, two particular histograms are recorded at the pump wavelengths of 1029.16 and 1030.90 nm as denoted by the dashed lines in Fig. \ref{fig3}(b). A temporal separation of 1.046 ns is manifested in the zooming-in illustration in Fig. \ref{fig5}(b), which is expected from the center-wavelength shift of 1.024 nm for the SFG spectrum around 790 nm. Note that all the histograms are located within the same range, which is defined by the bandwidth of the filter onto the upconversion spectrum. The observation of the slight histogram translation is made possible by the narrow pump bandwidth and precise time sampling. 

\begin{figure}[t!]
\includegraphics[width=0.8\columnwidth]{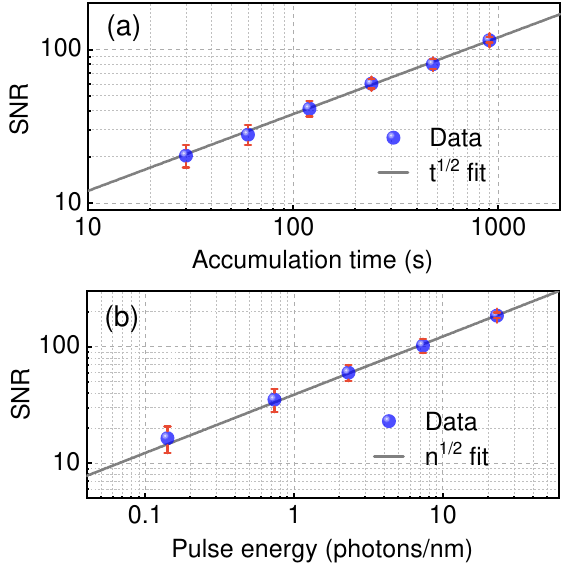}
\caption{MIR single-photon spectroscopy. (a) Signal-to-noise ratio (SNR) varies as the increase of accumulation time in the cace of an incident photon flux at 10 photons/nm/pulse. (b) SNR as a function of the pulse energy at a fixed integration time of 15 mins. Note that the fitted lines are given by the square root of the variables. The error bars denote standard deviations for five repetitive experiment measurements.}
\label{fig6}
\end{figure}

Furthermore, the ratio of the histogram at $t_2$ and the reference generates the time-stretch spectrum, which is used to reconstruct the SFG spectrum with the time-to-wavelength conversion factor. By taking into account the energy conservation during the upconversion process, the MIR absorption spectrum can be inferred as shown in Fig. \ref{fig5}(c), which agrees well with the Fourier transform infrared spectroscopy (FTIR) measurement at a resolution of 0.5 cm$^{-1}$. For our experimental settings, the maximum number of spectrally resolvable elements can in principle reach to 1775, which is inferred by the ratio of the pulse interval and temporal sampling resolution. The presented number of resolving channels is much larger than that for conventional dispersive spectrometers based on array detectors \cite{Zheng2023LPR}.

\begin{table*}[t]
\renewcommand\arraystretch{2}
\setlength{\tabcolsep}{21pt}
\caption{Performance comparison of broadband MIR single-photon spectrometers.}
\label{tab1}
\begin{tabular*}{0.9\linewidth}{@{}ccccc@{}}
\hline
 Ref. & Scheme & Range ($\mu$m) & Resolution (cm$^{-1}$) & Sensitivity \\ 
 \hline
 This & Time-stretch & 2.4-4.2 & 0.5 & 0.14 photons/nm/pulse \\ 
\cite{Zheng2023LPR} & SFG & 2.4-4.2 & 5 & 0.2 photons/nm/pulse \\
\cite{Dam2012NP} & SFG & \quad 2.84-3.1$^{[a]}$ & $\textgreater$6 & $\sim 10^{4}$ photons/second \\
\cite{Cai2022PR} & \quad SFG$^{[b]}$ & 3.1-3.8 & 10.5 & 0.09 photons/pulse \\ 
\cite{Kalashnikov2016NP} & SPDC & 4.1-4.5 & $\sim$27 & \quad /$^{[c]}$ \\  
\cite{Lindner21OE} & SPDC & 3.1-4.0 & \quad 0.56$^{[d]}$ & 11.9 photons/nm/pulse \\ 
\cite{Cheng2019NC} & SNSPD & \quad 0.6-2.0$^{[e]}$ & 26.6 &  \quad /$^{[c]}$ \\ 
\cite  {Kong2021NL} & SNSPD & 0.66-1.9 & 25.8 & \quad /$^{[c]}$ \\ 
\cite {Xiao2022ACS}& SNSPD & 1.2-1.7 & 23.8 & $\sim 10^{5}$ photons/second \\ 
\hline
\end{tabular*}
\begin{tablenotes}
\item[] $^{[a]}$It is possible cover a spectral range of 2.85-5 $\mu$m by tuning the crystal temperature.
\item[] $^{[b]}$The spectral measurement relies on a monochromator, which requires a grating scanning.
\item[] $^{[c]}$The spectrometer has a single-photon sensitivity, but the incident power is not specified.
\item[] $^{[d]}$The spectrometer is based on FTIR technique, which needs a mechanical delay scanning.
\item[] $^{[e]}$A higher resolution of 10.7 cm$^{-1}$ is possible with a reduced spectral coverage of 1.42-1.64 $\mu$m.
\end{tablenotes}
\end{table*}

Finally, we characterize the detection sensitivity of the presented MIR time-stretch spectrometer. In the low-light scenario, the MIR power of 1 mW is attenuated by 47 dB via a series of neutral density filters. The resultant illumination intensity on the sample is about 0.23 fJ/pulse, which corresponds to an incident photon flux of 10 photons/nm/pulse. The signal-to-noise ratio (SNR) is defined by the absorption peak at 2850 cm$^{-1}$ over the standard deviation of the background. Figure \ref{fig6}(a) presents the SNR as a function of the accumulation time during the histogram acquisition. The SNR increases with a longer accumulation time. At a long integration time of 900 s, the detection sensitivity can be improved to 0.14 photons/nm/pulse, as given in Fig. \ref{fig6}(b). The SNR shows a square-root dependence on the integration time or input pulse energy. Note that the incident power is determined as the illumination power on the sample. Actually, the detected photon number on the APD is much smaller, which is estimated to be 2.7$\times10^{-5}$ photons/pulse from the division between the registered event in the histogram and the total pulse number within the accumulation time. In the coincidence counting configuration, only signals phase locked to the trigger signal in TCSPC can be effectively extracted, which helps to suppress the severe random background noise in photon-starving scenarios \cite{Yang2022SB}.

\section{Discussion and conclusion}
Over the past decades, the frequency upconversion technique has emerged as a promising approach to circumvent the limitations in the current infrared detector technology, which is widely used in sensitive detection, imaging and spectroscopy in the MIR region \cite{Barh2019AOP}. In previous reports, broadband MIR upconversion spectrometers are commonly implemented in the spatial domain, which are based on angularly dispersive elements and multi-pixel array sensors. In this configuration, the number of spectrally resolvable elements is dictated by the pixel number of the detector, which imposes a trade-off between spectral coverage and spectral resolution \cite{Zheng2023LPR, Wolf2017OE, Rodrigo2021LPR}. The performance compromise becomes more prominent in the case of using single-photon array detectors, since high-pixel-density spatial formats are hard to access \cite{Zheng2023LPR}. It thus remains a long-standing quest to realize MIR single-photon spectroscopy at high resolution over a broad spectral range. 

Here, we leverage the wavelength-time spectroscopy to mitigate the resolution-bandwidth trade-off, where high-sensitivity and high-resolution measurements are performed in the time-domain via a single-element single-photon detector. The presented MIR time-stretch spectrometer features single-pixel simplicity and high optical throughput, which contrasts with the monochromator-based scheme that relies on mechanical grating scanning and stringent spectral filtering \cite{Cai2022PR}. In our experiment, the involved synchronous pumping architecture is the key to implementing the high-performance time-stretch upconversion spectrometer. First, the chirped pulsed pump allows a high-fidelity nonlinear upconversion in a high-efficiency and low-noise fashion \cite{Zheng2023LPR}, thus resulting in high sensitivity and high resolution in subsequent spectroscopic measurements. Second, the upconverted field is located in the visible or near-infrared regions, where high-bandwidth detectors and low-loss dispersive media are available to implement the high-precision temporal sampling \cite{Goda2013NP}. Third, the stable and precise synchronization source facilitates the use of TCSPC technique to reconstruct the time-stretch waveforms at the single-photon level \cite{Davis2017OE, Yang2022SB}. Fourth, the photon energy of the pump field is lower than that for the upconverted field, $\textit{i.e.}$, $\hbar \omega_\text{c} =  \hbar \omega_\text{p} + \hbar \omega_\text{s}$, where $p,s,c$ denote the pump, signal, and converted field. The underlying SFG process enables a noise-free coherent conversion process. In contrast, previous demonstrations for upconversion time-stretch spectroscopy rely on the DFG process with $\hbar \omega_\text{c} =  \hbar \omega_\text{p} - \hbar \omega_\text{s}$ \cite{Wen2023LPR, Hashimoto2023LSA}, which suffer from severe noises from SPDC fluorescence under intensive pumping. The intrinsically noisy conversion impedes the reported instantiations from the single-photon operation.

Table \ref{tab1} summarizes representative demonstrations on the broadband MIR single-photon spectroscopy. In comparison to conventional SFG-based schemes \cite{Zheng2023LPR, Dam2012NP, Cai2022PR}, our infrared spectrometer provides over tenfold improvement on the spectral resolution. Another promising approach is based on the quantum nonlinear interferometry of correlated photons from non-degenerate SPDC, which typically requires spatially resolved detection of the interference pattern with a camera \cite{Kalashnikov2016NP}. Recently, the use of FTIR technique allows for an improved spectral resolution of 0.56 cm$^{-1}$, albeit with the requirement of delay scanning operation for the data acquisition \cite{Lindner21OE}. In addition, the superconducting nanowire single-photon detectors (SNSPDs) are integrated to realize on-chip miniature spectrometers, yet facing challenges in longer-wavelength extension and resolution enhancement \cite{Cheng2019NC, Kong2021NL, Xiao2022ACS}. Therefore, our presented work not only establishes an effective path to realizing ultra-sensitive MIR time-stretch spectroscopy, but also represents a landmark in high-resolution MIR single-photon spectrometers. 

In conclusion, we have implemented for the first time a MIR single-photon spectrometer in the time domain, which combines the merits from frequency upconversion detection and time-stretch spectroscopy. The presented infrared spectrometer overcomes the stringent requirement for high-definition array detectors in precise spectroscopic measurements, and allows for sub-wavenumber resolution over a wide spectral coverage based on a single-photon detector. The spectral resolution can be further enhanced by using a larger-GVD disperser and a lower-timing-jitter detector. Moreover, it is feasible to extend the presented approach into longer infrared or terahertz regions \cite{Rodrigo2021LPR}, where high-sensitivity and high-resolution spectroscopy is highly demanded. Notably, the MIR single-photon spectrometer features single-pixel simplicity and high optical throughput, which would provide a practical alternative or complementary technique for applications in materials analysis and chemical sensing.
\newline
\newline

\section*{Acknowledgements}
This work was supported by National Key Research and Development Program (2021YFB2801100), National Natural Science Foundation of China (62175064, 62235019, 62035005, 12022411); Shanghai Pilot Program for Basic Research (TQ20220104); Natural Science Foundation of Chongqing (CSTB2023NSCQ-JQX0011, CSTB2022NSCQ-MSX0451, CSTB2022NSCQ-JQX0016); Shanghai Municipal Science and Technology Major Project (2019SHZDZX01); Fundamental Research Funds for the Central Universities.

\section*{Conflict of Interest}
The authors declare no conflict of interests.

\section*{Data Availability Statement}
The data that support the findings of this study are available from the corresponding author upon reasonable request.

\section*{Keywords}
mid-infrared spectroscopy, single-photon spectrometer, frequency upconversion, time-stretch spectroscopy

\end{document}